\documentclass[aps,prc,twocolumn,superscriptaddress,showpacs]{revtex4}

\usepackage{graphicx}
\usepackage{dcolumn}
\usepackage{bm}
\usepackage{ulem}
\usepackage{color}

\newcommand{\al}{$\alpha$}

\newcommand{\rag}{($\alpha$,$\gamma$)}

\newcommand{\rap}{($\alpha$,$p$)}
\newcommand{\rpa}{($p$,$\alpha$)}

\newcommand{\rpt}{($p$,$t$)}

\newcommand{\Nsv}{$N_A < \sigma v >$}

\newcommand{\rppro}{$rp$-process}

\newcommand{\sfact}{S-factor}

\begin{document}

\title{
  Examination and experimental constraints of the stellar reaction rate factor
  $N_A < \sigma v >$ of the $^{18}$Ne($\alpha$,$p$)$^{21}$Na reaction at
  temperatures of X-Ray Bursts
}

\author{P.\ Mohr}
\email[Email: ]{WidmaierMohr@t-online.de}
\affiliation{
Diakonie-Klinikum, D-74523 Schw\"abisch Hall, Germany}
\affiliation{
Institute of Nuclear Research (ATOMKI), H-4001 Debrecen, Hungary}

\author{A.\ Matic}
\altaffiliation[Present address: ]{IBA Particle Therapy, D-45157 Essen, Germany}
\affiliation{Kernfysisch Versneller Instituut, University of Groningen,
Zernikelaan 25, NL-9747 AA Groningen, The Netherlands}

\date{\today}

\begin{abstract}
The $^{18}$Ne($\alpha$,$p$)$^{21}$Na reaction is one key for the break-out from
the hot CNO-cycles to the $rp$-process. Recent papers have provided reaction
rate factors $N_A < \sigma v >$ 
which are discrepant by at least one order of
magnitude. The compatibility of the latest experimental results is tested,
and a partial explanation for the discrepant $N_A<\sigma v>$ is given. A new
rate factor is derived from the combined analysis of all available data. The new
rate factor is located slightly below the higher rate factor by Matic {\it et
  al.}\ at low temperatures and significantly below at higher temperatures
whereas it is about a factor of five higher than the lower rate factor
recently published by Salter {\it et al.}
\end{abstract}


\maketitle

\section{Introduction}
\label{sec:intro}
The reaction rate of the $^{18}$Ne\rap $^{21}$Na reaction provides a route
from hot CNO-cycles to the NeNa and MgAl cycles and finally to the
\rppro\ at typical temperatures of e.g.\ about $1-2$\,GK 
($T_9 = 1 - 2$) in X-ray bursters (XRB) \cite{Sch06}. It is expected that
this reaction is the dominating route in the low temperature range
\cite{Wie99}. An alternative route from hot CNO-cycles to the \rppro\ may
be the $^{15}$O\rag $^{19}$Ne reaction. 

The relatively high temperatures correspond to most effective
energies of about 1.3 to 2.1\,MeV for the $^{18}$Ne\rap $^{21}$Na reaction
which are experimentally well accessible. However, experiments remain very
difficult because of the short-living $^{18}$Ne nucleus ($T_{1/2} = 1.67$\,s)
and the
limited intensity of radioactive beam facilities. Thus, besides the direct
approach of measuring the $^{18}$Ne\rap $^{21}$Na reaction cross section
\cite{Bra99,Gro02}, the reverse $^{21}$Na\rpa $^{18}$Ne
reaction has been studied very recently \cite{Sal12} and in an earlier
unpublished experiment \cite{ANL},
and the resonance energies have been determined from various
transfer experiments populating states in the compound $^{22}$Mg nucleus
\cite{Chen01,Cagg02,Berg03,Chae09,Matic09}. 

The focus of
the present paper is the comparison of the latest experiments by Groombridge
{\it et al.}\ (hereafter: GRO) \cite{Gro02}, Salter {\it et al.}\ (SAL)
\cite{Sal12}, Chae {\it et al.}\ (CHA) \cite{Chae09},
and Matic {\it et al.}\ (MAT) \cite{Matic09}. The earlier
direct data of \cite{Bra99} have been improved and extended by the same group
leading to the GRO data. The SAL data are the only published data for the
inverse reaction; a brief comparison to the unpublished data measured at
Argonne National Laboratory (ANL) is also provided. 
The MAT transfer data have by far the best energy resolution
which is essential for a precise determination of the resonance
energies. Additional measurements of angular distributions in CHA
lead to a new spin assignment only in few cases (see Table III of CHA).

The reaction rate factor \Nsv\ 
for the $^{18}$Ne\rap $^{21}$Na reaction is given by
the sum over the contributing resonances:
\begin{eqnarray}
\frac{N_A < \sigma v >}{ {\rm{cm}}^3 \, {\rm{s}}^{-1} \, {\rm{mol}}^{-1} }
\nonumber 
& = &
1.54 \times 10^{11} \, (\mu \, T_9)^{-3/2} \\
& \times &
\sum_i (\omega \gamma)_i \times \exp{(-11.605 E_i/T_9)}
\label{eq:rate}
\end{eqnarray}
with the reduced mass $\mu$ in units of amu, the resonance energies
$E_i$ in MeV, and the resonance strengths $(\omega \gamma)_i$ in MeV. In
general, resonance energies are given as $E$ in the center-of-mass (c.m.)\
system without index; excitation energies are given as $E^\ast$ in 
this paper. 

The resonance strength $(\omega \gamma)$ for the $^{18}$Ne\rap
$^{21}$Na reaction is given by
\begin{equation}
\omega \gamma = (2J+1) \, \frac{\Gamma_\alpha \Gamma_p}{\Gamma}
\label{eq:strength}
\end{equation}
with the resonance spin $J$, the partial widths $\Gamma_\alpha$ and $\Gamma_p$,
and the total width $\Gamma = \Gamma_\alpha + \Gamma_p + \Gamma_\gamma \approx
\Gamma_\alpha + \Gamma_p$. In most cases it can be expected that
$\Gamma_\alpha \ll \Gamma_p$, and thus $(\omega \gamma) \approx (2J+1) \,
\Gamma_\alpha$. The application of the simple formula 
for narrow resonances in Eq.~(\ref{eq:rate})
is justified because the resonance widths $\Gamma$ are much smaller than the
resonance energies $E$ \cite{Matic07}.

In the following we first briefly review the various experimental approaches
and discuss the resulting uncertainties in the determination of the reaction
rate factor \Nsv . Next we check whether the experimental results of GRO, MAT,
CHA, 
SAL, and ANL are compatible with each other. Finally, the reaction rate factors
\Nsv\ of the different studies are compared. Note that \Nsv\ of different
studies may differ not only from discrepant resonance energies and resonance
strengths or cross sections, but also from a different number of considered
resonances in Eq.~(\ref{eq:rate}).

\section{Brief review of available data}
\label{sec:rev}
\subsection{Transfer data}
\label{sec:trans}
Various transfer experiments have been performed in the last decade to study
properties of the compound nucleus $^{22}$Mg
\cite{Chen01,Cagg02,Berg03,Chae09,Matic09}. A detailed comparison of the
results is provided in MAT and is not repeated. Here we briefly summarize the
MAT results and some modifications resulting from the CHA data.

Transfer data are able to provide excitation energies $E^\ast$ and spin and
parity $J^\pi$ of states in $^{22}$Mg. However, from the transfer data it is
not possible to determine resonance strengths $\omega \gamma$ which are
the second ingredient for the calculation of the rate factor \Nsv\ in
Eq.~(\ref{eq:rate}).

\subsubsection{Matic {\it et al.}}
\label{sec:mat}
In the MAT approach the $^{24}$Mg\rpt $^{22}$Mg reaction is used to populate
excited states in the compound nucleus $^{22}$Mg at proton energies of
slightly below 100\,MeV. The experiment has been performed using the Grand
Raiden spectrometer at RCNP, Osaka. The excellent energy
resolution of about 13\,keV allows a precise determination of excitation
energies $E^\ast$ which enter exponentially into \Nsv\ in Eq.~(\ref{eq:rate})
via $E^* = E + S_\alpha$ (with the separation energy of the \al\ particle in
$^{22}$Mg of $S_\alpha = 8.142$\,MeV) and are thus the
main source of uncertainties. ($S_\alpha = 8.142$\,MeV is taken from the new
Audi and Meng compilation \cite{AMDC}; the small difference to the earlier
result of $S_\alpha = 8.139$\,MeV \cite{Audi03} does practically not affect
the rate factor \Nsv\ in the relevant temperature range around $T_9 = 1 - 2$.)

In addition to the excitation energies $E^\ast$, the total widths $\Gamma$ can
be determined from these data by fitting the observed peak widths
$\Gamma_{\rm{obs}}$. Most of the observed states are much broader than the
experimental resolution, and thus the required unfolding procedure leads only
to minor additional uncertainties for the derived width $\Gamma$. The results
are listed in Table \ref{tab:width}. As can be seen from Table \ref{tab:width},
practically all resonances fulfill the criterion of $\Gamma/E \le 0.1$ which
is often used as definition for narrow resonances (although also more
stringent definitions for narrow resonances can be found in literature).
As we will show in Sect.~\ref{sec:CHA}, the simple formula for narrow
resonances in Eq.~(\ref{eq:rate}) provides the reaction rate factor \Nsv\ for
the $^{18}$Ne($\alpha$,$p$)$^{21}$Na reaction with sufficient accuracy. In
this sense the resonances in Table \ref{tab:width} can be considered generally
as narrow resonances.
\begin{table}[tbh]
\caption{\label{tab:width}
Excitation energy $E^\ast$, resonance energy $E$, spin and parity $J^\pi$, 
total width $\Gamma$, and resonance strength $\omega \gamma$
for excited states in $^{22}$Mg from the $^{24}$Mg\rpt $^{22}$Mg experiment in
\cite{Matic09}. Later revisions for individual states are marked by
``$^{\ast {\rm{GRO}}}$'' and ``$^{\ast {\rm{CHA}}}$''; these revisions are based
on the replacement of the experimental resonance strengths of GRO by
calculated resonance strengths and on
revised (but still tentative) spin assignments by CHA \cite{Chae09} (see also
Table \ref{tab:mod}). The finally recommended strengths will be slightly lower
by a factor of 0.55 (see discussion in Sect.~\ref{sec:MATSAL} and
\ref{sec:rate}).
}
\begin{tabular}{ccrr@{$\pm$}lr@{$\times$}l}
\hline
\multicolumn{1}{c}{$E^\ast$ (MeV)}
& \multicolumn{1}{c}{$E$ (MeV)}
& \multicolumn{1}{c}{$J^\pi$}
& \multicolumn{2}{c}{$\Gamma$ (keV)}
& \multicolumn{2}{c}{$\omega \gamma$ (eV)} \\
\hline
  8.182 & 0.040 & $[2^+]$ &  33.5  &    2.2 & 8.53 & $10^{-65}$ \\
  8.385 & 0.243 & $[2^+]$ &  47.0  &    5.3 & 1.33 & $10^{-17}$ \\
  8.519 & 0.377 & $[3^-]$ &  25.7  &    4.1 & 4.87 & $10^{-14}$ $^{\ast {\rm{CHA}}}$ \\
  8.574 & 0.432 & $[4^+]$ &  20.6  &   16.8 & 3.26 & $10^{-12}$ \\
  8.657 & 0.515 & $[0^+]$ &  15.5  &    3.5 & 4.97 & $10^{-8 }$ \\
  8.743 & 0.601 & $[4^+]$ &  65.5  &   22.8 & 5.15 & $10^{-9 }$ \\
  8.783 & 0.641 & $[1^-]$ &  22.5  &    7.0 & 1.21 & $10^{-5 }$ \\
  8.932 & 0.790 & $[2^+]$ &  51.6  &    5.9 & 4.13 & $10^{-4 }$ \\
  9.080 & 0.938 & $[1^-]$ & 114.4  &   19.7 & 2.31 & $10^{-2 }$ \\
  9.157 & 1.015 & $[4^+]$ & \multicolumn{2}{c}{$< 20.5$} & 8.70 & $10^{-4 }$ \\
  9.318 & 1.176 & $[2^+]$ &  22.6  &    8.0 & 4.97 & $10^{-1 }$ \\
  9.482 & 1.340 & $[3^-]$ & \multicolumn{2}{c}{$< 6.3$} & 1.25 & $10^{-1 }$ \\
  9.542 & 1.400 & $[2^+]$ & \multicolumn{2}{c}{$< 22.9$} & 1.78 & $10^{0  }$ $^{\ast {\rm{CHA}}}$ \\
  9.709 & 1.567 & $[0^+]$ & 267.8  &   48.2 & 5.18 & $10^{1  }$ \\
  9.752 & 1.610 & $[2^+]$ &  31.4  &    6.8 & 8.22 & $10^{0  }$ $^{\ast {\rm{CHA}}}$ \\
  9.860 & 1.718 & $[0^+]$ & 121.3  &   10.4 & 2.07 & $10^{1  }$ \\
 10.085 & 1.943 & $[2^+]$ &  25.8  &    9.3 & 2.25 & $10^{2  }$ \\
 10.272 & 2.130 & $2^+$   &  20.7  &    2.7 & 1.03 & $10^{4  }$ $^{\ast {\rm{GRO}}}$ \\
 10.429 & 2.287 & $[4^+]$ & 144.2  &   25.8 & 7.30 & $10^{3  }$ $^{\ast {\rm{GRO}}}$ \\
 10.651 & 2.509 & $[3^-]$ &  72.8  &   19.1 & 1.82 & $10^{4  }$ $^{\ast {\rm{GRO}}}$ \\
 10.768 & 2.626 & $[2^+]$ &  94.9  &   29.6 & 1.16 & $10^{4  }$ \\
 10.873 & 2.731 & $[0^+]$ &  40.2  &   12.0 & 4.52 & $10^{4  }$ $^{\ast {\rm{GRO}}}$ \\
 11.001 & 2.859 & $[4^+]$ & 135.8  &   12.9 & 8.10 & $10^{3  }$ $^{\ast {\rm{GRO}}}$ \\
 11.315 & 3.173 & $[4^+]$ & 203.7  &   37.0 & 1.83 & $10^{3  }$ \\
 11.499 & 3.357 & $[2^+]$ & 116.8  &   21.8 & 8.64 & $10^{4  }$ \\
 11.595 & 3.453 & $[1^-]$ &  48.3  &   14.7 & 6.11 & $10^{4  }$ $^{\ast {\rm{CHA}}}$ \\
 11.747 & 3.605 & $[0^+]$ & 166.1  &   64.4 & 7.13 & $10^{4  }$ \\
 11.914 & 3.772 & $[0^+]$ & 122.4  &   19.7 & 8.82 & $10^{4  }$ $^{\ast {\rm{CHA}}}$ \\
 12.003 & 3.861 & $[1^-]$ 
  & \multicolumn{2}{c}{$-$} \footnote{state adopted from \cite{Berg03}}
  & 4.31 & $10^{5  }$ \\
 12.185 & 4.043 & $[3^-]$ & 236.4  &   52.0 & 2.60 & $10^{5  }$ \\
 12.474 & 4.332 & $[2^+]$ & 193.8  &   51.6 & 3.89 & $10^{5  }$ \\
 12.665 & 4.523 & $[3^-]$ & 128.8  &   23.5 & 3.45 & $10^{5  }$ \\
 13.010 & 4.868 & $[0^+]$ & 600.9  &  114.5 & 2.16 & $10^{5  }$ \\
\hline
\end{tabular}
%
\end{table}

The resonance strengths $\omega \gamma$ in MAT have generally
been calculated by the
following procedure. In a first step spin and parity $J^\pi$ for the states
seen in the $^{24}$Mg\rpt $^{22}$Mg experiment were tentatively
assigned from the mirror nucleus $^{22}$Ne. Note that the level scheme for the
stable mirror nucleus $^{22}$Ne is well-established up to relatively high
excitation energies. In some cases also theoretical predictions from the shell
model have been used \cite{Brown88}.
Next, upper limits for $\Gamma_\alpha$ were
calculated from the single-particle Wigner limit using the tentative spin
assignments (see Table VII in MAT, partly repeated in Table \ref{tab:width} in
this work). Finally, the Wigner limit has been scaled by carefully chosen
reduced widths. Whenever possible, the reduced \al -widths were taken from
experimental data obtained for the stable mirror nucleus $^{22}$Ne by \al
-transfer on the mirror target $^{18}$O. In the remaining cases where no
experimental information is available, the reduced \al -widths were estimated
by simple but reasonable theoretical assumptions. A detailed discussion is
given in Sect.~V in MAT.

In the few cases where experimental data are available from the
GRO study, experimental resonance strengths were used by MAT in their
calculation of \Nsv . Levels from threshold up to an excitation energy of 
about 13\,MeV (corresponding to $E \approx 5$\,MeV) are taken into account in
MAT (see their Table VII), and thus the resulting rate factor \Nsv\ is well
determined over a broad temperature range including the astrophysically most
relevant range of $T_9 = 1 - 2$. 

For an independent comparison of the various
studies of the $^{18}$Ne\rap $^{21}$Na reaction, we replace the experimental
resonance strengths from the GRO data by calculations similar to the other
resonance strengths in MAT (see upper part of Table \ref{tab:mod}; these data
will be referenced as MAT-th: ``th'' for theoretical strengths only). A further
explanation for this replacement will become visible later in the comparison
of the MAT data to the GRO data and SAL data (see Sect.~\ref{sec:GROSAL}).
\begin{table}[tbh]
\caption{\label{tab:mod}
Revisions for resonance strengths $\omega \gamma$, compared to Table VII in
MAT and Table \ref{tab:width}. The upper five lines are recalculations of
$\omega \gamma$ instead of the resonance strengths adopted from GRO
(here $\omega \gamma_{\rm{MAT}}$ is identical to $\omega \gamma_{\rm{GRO}}$). 
The lower five lines result from new spin 
assignments in CHA. For comparison, the original strengths $\omega
\gamma_{\rm{MAT}}$ from MAT are also listed.
}
\begin{tabular}{rrcr@{$\times$}lr@{$\times$}lc}
\hline
\multicolumn{1}{c}{$E^\ast$ (MeV)}
& \multicolumn{1}{c}{$E$ (MeV)}
& \multicolumn{1}{c}{$J^\pi$}
& \multicolumn{2}{c}{$\omega \gamma$ (eV)}
& \multicolumn{2}{c}{$\omega \gamma_{\rm{GRO}}$ (eV)}
& \multicolumn{1}{c}{} \\
\hline
 10.272 & 2.130 & $2^+$    & 1.31 & $10^3$       & 1.03 & $10^4$ \\
 10.429 & 2.287 & $[4^+]$  & 4.89 & $10^1$       & 7.30 & $10^3$ \\
 10.651 & 2.509 & $[3^-]$  & 1.12 & $10^3$       & 1.82 & $10^4$ \\
 10.873 & 2.731 & $[0^+]$  & 1.19 & $10^4$       & 4.52 & $10^4$ \\
 11.001 & 2.859 & $[4^+]$  & 5.81 & $10^2$       & 8.10 & $10^3$ \\
\hline
\multicolumn{1}{c}{$E^\ast$ (MeV)}
& \multicolumn{1}{c}{$E$ (MeV)}
& \multicolumn{1}{c}{$J^\pi$}
& \multicolumn{2}{c}{$\omega \gamma$ (eV)}
& \multicolumn{2}{c}{$\omega \gamma_{\rm{MAT}}$ (eV)}
& \multicolumn{1}{c}{$J^\pi_{\rm{MAT}}$} \\
\hline
  8.519 & 0.377 & $[2^+]$  & 1.53 & $10^{-11}$  & 4.87 & $10^{-14}$ & $[3^-]$ \\
  9.542 & 1.400 & $[1^-]$  & 1.31 & $10^{1}$    & 1.78 & $10^{0}$   & $[2^+]$ \\
  9.752 & 1.610 & $[1^-]$  & 4.82 & $10^{1}$      & 8.22 & $10^{0}$ & $[2^+]$ \\
 11.595 & 3.453 & $[4^+]$  & 3.67 & $10^{3}$      & 6.11 & $10^{4}$ & $[1^-]$ \\
 11.914 & 3.772 & $[2^+]$  & 1.77 & $10^{5}$      & 8.82 & $10^{4}$ & $[0^+]$ \\
\hline
\end{tabular}
%
\end{table}

\begin{figure}[htb]
\includegraphics[width=\columnwidth,clip=]{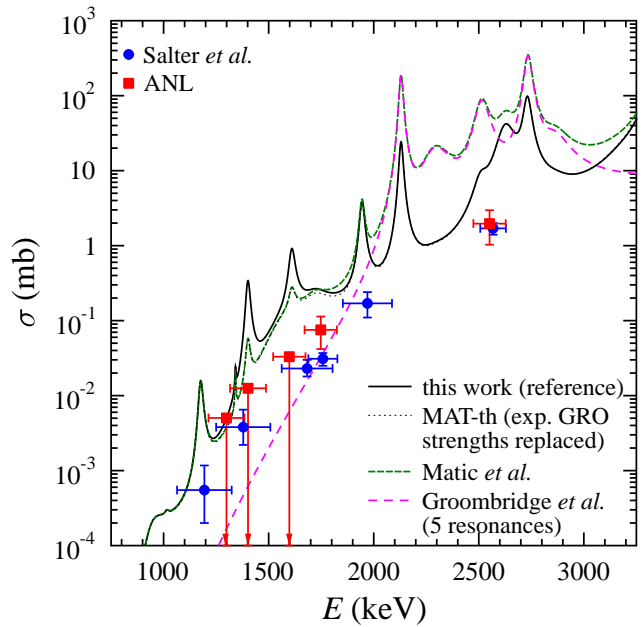}
\caption{
\label{fig:sigma}
(Color online)
Reaction cross section of the $^{18}$Ne\rap $^{21}$Na reaction, calculated
from total widths $\Gamma$ (see Table \ref{tab:width}) and from resonance
strengths $\omega \gamma$. The full black line uses the MAT strengths with
all revisions in Table \ref{tab:mod} (reference cross section
$\sigma_{\rm{ref.}}$). The green short-dashed line refers to 
the original MAT strengths (from their Table VII). The long-dashed magenta
line shows the five GRO resonances only. 
The black dotted line (MAT-th) uses the MAT strengths but replaces the
experimental GRO strengths by the calculated strengths (see first five lines
in Table \ref{tab:mod}). MAT-th is almost identical to MAT (green
short-dashed) at low energies and to $\sigma_{\rm{ref.}}$ (full black) at
higher energies; thus, the black dotted MAT-th curve becomes only visible at
energies around $E \approx 1.8$\,MeV.
The experimental data points have been determined by the
inverse $^{21}$Na\rpa $^{18}$Ne reaction and represent the ground state
contribution only. Further discussion see text.
}
\end{figure}

Under the realistic assumption that the total width $\Gamma$ is dominated by
the proton partial width $\Gamma_p$ (and thus $\omega \gamma \approx \omega
\Gamma_\alpha$ and $\Gamma \approx \Gamma_p$), 
the reaction cross section $\sigma(E)$ as a function of energy
can be calculated as a sum over Breit-Wigner resonances: $\sigma(E) 
= \sum_i \sigma_{BW,i}(E)$ with
\begin{equation}
\sigma_{BW}(E) = \frac{\pi \hbar^2}{2 \mu E} \, 
\frac{\omega \Gamma_\alpha \Gamma_p}{(E-E_R)^2 + \Gamma^2/4} \quad \quad .
\label{eq:BW}
\end{equation}
The result is shown
in Fig.~\ref{fig:sigma} (MAT-th, thin dotted black line). The upper limit of
the total width $\Gamma$ has been used for the three resonances at $E^\ast =$
9.157, 9.482, and 9.542\,MeV. The resulting uncertainty from these upper
limits for the reaction rate factor \Nsv\ remains negligible because the
resonances are narrow in any case.

This calculation of the cross section $\sigma(E)$ enables a
more detailed comparison of the data from transfer and from the study of the
inverse reaction (see Sect.~\ref{sec:MATSAL}). The original strengths of MAT
(using the experimental GRO strengths where available)
lead to the green short-dashed curve in Fig.~\ref{fig:sigma} which is much
higher at energies between 2 and 3\,MeV because of the high resonance
strengths taken from the GRO data.

\subsubsection{Chae {\it et al.}}
\label{sec:CHA}
The CHA data are also based on a study of the $^{24}$Mg\rpt $^{22}$Mg
reaction. The experiment has been performed at lower energies slightly above
40\,MeV at the Holifield Radioactive Ion Beam Facility at Oak Ridge. The
achieved energy resolution does not reach the MAT data, but the larger angular
range under study allows an improved but still tentative assignment of spin
and parity $J^\pi$ of several states in $^{22}$Mg. 

We have taken the tentative assignments from the CHA data and have
recalculated the resonance strengths $\omega \gamma$ for these states using
the adopted energies of the high-resolution MAT data. The results are listed
in Table \ref{tab:mod} (lower part). The new spin assignments lead to smaller
spins $J$ for some low-lying resonances and consequently to larger resonance
strengths $\omega \gamma$ because of the reduced centrifugal barrier. However,
the rate factor \Nsv\ is given as the sum over many resonances, and the
influence of the new spin assignments on the total rate factor remains below a
factor of two over the astrophysically relevant temperature range of $T_9 = 1
- 2$.

The calculated resonance strengths $\omega \gamma$ from Table VII in MAT with
all modifications from Table \ref{tab:mod} are taken as a reference in this
work to calculate $\sigma_{\rm{ref.}}(E)$ and the rate factor \Nsv
$_{\rm{ref.}}$. For simplicity, the rate factor \Nsv\ has been calculated from
the approximation in Eq.~(\ref{eq:rate}). Thus, it will also be possible to
modify our rate factor without much efforts as soon as improved experimental
data for any of the $^{18}$Ne\rap $^{21}$Na resonances will become available.
The result is shown as the full black line in Fig.~\ref{fig:sigma}. Note that
the final recommendation for $\sigma(E)$ and \Nsv\ from the combined analysis
of all available experimental data will be almost a factor of 2 lower (see
Sect.~\ref{sec:rate}). 

We have tested the validity of the simple rate formula for narrow resonances in
Eq.~(\ref{eq:rate}) by a numerical integration of the $\sigma_{\rm{ref.}}(E)$
curve, and it is found that the deviation between the numerical integration
and the simple formula in Eq.~(\ref{eq:rate}) remains below 10\,\% over the
full temperature range under study ($T_9 = 0.25 - 3$) and far below 5\,\% for
the most relevant range of $T_9 = 1 - 2$.

It is obvious that the calculated cross section $\sigma(E)$ and reaction rate
factor \Nsv\ depend on the spin and parity $J^\pi$ of the considered
resonances. A detailed study of the corresponding uncertainties has already
been given in MAT using random spin
assignments, and it was concluded that \Nsv\ does not change by more than one
order of magnitude. Here we provide reduction factors
$\Gamma_{s.p.}(L=0)/\Gamma_{s.p.}(L)$ for angular momenta $0 \le L \le 4$ in the
energy range between 1 and 4\,MeV (see Table \ref{tab:red}) where the
\al\ single-particle widths $\Gamma_{s.p.}$ have been calculated similar to
MAT. The largest reduction factors are found for large angular momenta $L$ at
low energies where the influence of the centrifugal barrier is most
important. Here an increase of $L$ by 1 leads to a reduction of a factor of 10
whereas at higher energies and small $L$ the increase of 1 leads to a
reduction of less than a factor of 2.
\begin{table}[tbh]
\caption{\label{tab:red}
Reduction factors $\Gamma_{s.p.}(L=0)/\Gamma_{s.p.}(L)$ for angular momenta $0
\le L \le 4$ in the energy range between 1 and 4\,MeV.
}
\begin{tabular}{crrrrr}
\hline
\multicolumn{1}{c}{$E$ (MeV)}
& \multicolumn{1}{c}{$L=0$}
& \multicolumn{1}{c}{$L=1$}
& \multicolumn{1}{c}{$L=2$}
& \multicolumn{1}{c}{$L=3$}
& \multicolumn{1}{c}{$L=4$} \\
\hline
 1.0 & $\equiv 1.0$ &    1.9 &     7.0 &     44.4 &    463.1 \\
 2.0 & $\equiv 1.0$ &    1.7 &     5.1 &     23.8 &    171.1 \\
 3.0 & $\equiv 1.0$ &    1.5 &     3.4 &     11.8 &     60.9 \\
 4.0 & $\equiv 1.0$ &    1.3 &     2.2 &      5.6 &     20.9 \\
\hline
\end{tabular}
%
\end{table}

The assignment of spin and parity $J^\pi$ of the states in $^{22}$Mg (e.g.\ in
MAT) is tentative and has mainly been made from the mirror nucleus
$^{22}$Ne. Even if some of the individual assignments may be incorrect, the
distribution of spins should be correct because the $J^\pi$ assignments in
$^{22}$Ne are mainly firm and not tentative. The cross section $\sigma(E)$ of
the $^{18}$Ne\rap $^{21}$Na reaction is composed of overlapping resonances,
and the calculation of the rate factor \Nsv\ requires folding with the
Maxwell-Boltzmann distribution. Thus, at any astrophysically relevant
temperature $T_9 > 1$ the rate factor is defined by a sum over several
contributing resonances, and the influence of a modified spin assignment
of an individual resonance remains very limited, in particular, if the spin
distribution is kept; i.e., the modification of $J^\pi$ of one resonance
(e.g.\ increase of $L$ and reduced $\Gamma_\alpha$ and $\omega \gamma$) is
compensated by a similar modification of $J^\pi$ of another resonance
(decrease of $L$ and enhanced $\Gamma_\alpha$ and $\omega \gamma$).

As we will see later, the modified spin assignments of the CHA experiment
(mainly smaller $J^\pi$ than used by MAT, see Table \ref{tab:mod}, lower part)
will lead to a marginal enhancement of the rate factor \Nsv\ around $T_9
\approx 1$ by less than a factor of 2. In combination with the above
arguments, it seems thus reasonable that the uncertainty of the spin
assignments for the calculated rate factor \Nsv\ does not exceed a factor of
two in the full temperature range under study.

\subsection{Direct data}
\label{sec:direct}
The GRO experiment has been performed at the Radioactive Ion Beam facility at
Louvain-la-Neuve. They have measured the $^{18}$Ne\rap $^{21}$Mg reaction
directly using an extended $^4$He gas target and a $^{18}$Ne beam. The chosen
detection technique allowed the reconstruction of the interaction vertex
in the extended gas target, and together with the measured proton energy it
was possible to determine the energies and resonance strengths of 8 resonances
and their main decay branch into the $p_i$ channel ($p_0$ corresponds to the
$^{21}$Na ground state, $p_1$ to the first excited state, and so on). The data
cover the energy range from about 1.7 to 2.9\,MeV. The rate
factor \Nsv\ was calculated from these 8 resonances; this \Nsv\ is considered
as a lower limit by GRO because of the missing resonances outside the studied
energy region and because of perhaps missed weak resonances or missed weak
branches of observed resonances inside the studied energy region. 

The cross section $\sigma(E)$ as a function of energy is calculated from 5
adopted resonances (see discussion below in Sect.~\ref{sec:MATGRO}) using
excitation energies $E^\ast$ from MAT, total widths $\Gamma$ from Table
\ref{tab:width}, and resonance strengths $\omega \gamma$ from GRO. The result
is shown in Fig.~\ref{fig:sigma} with a long-dashed magenta line. Obviously,
in the energy region between 2 and 3\,MeV 
it is close to the MAT result because MAT have used resonance strengths from
GRO, but it is much higher than the reference calculation of this work.

Finally, it is interesting to note that the summed $p_0$ resonance strength to
the ground state is about 42\,\% of the total summed strength in the GRO
data. This number will be relevant for comparison with the SAL and ANL data
for the inverse $^{21}$Na\rpa $^{18}$Ne reaction.

\subsection{Data for the inverse reaction}
\label{sec:inv}
The latest experiment by SAL and the unpublished ANL experiment have used
the reverse $^{21}$Na\rpa $^{18}$Ne
reaction in inverse kinematics with a radioactive $^{21}$Na beam and a solid
CH$_2$ target. An average cross section $\bar{\sigma}$ at the energy
$E_{\rm{eff}}$ is determined from the measured \al\ yield:
\begin{equation}
\bar{\sigma}(E_{\rm{eff}}) = 
\frac{1}{2\, \Delta E} \,
\int_{E_{\rm{eff}}-\Delta E}^{E_{\rm{eff}}+\Delta E} \sigma(E) \, dE
\label{eq:av}
\end{equation}
where $E_{\rm{eff}}$ is the energy in the center of the CH$_2$ target, and the
total energy loss in the target is given by $2\, \Delta E$. In the energy
range under study the $^{21}$Na\rpa $^{18}$Ne cross section populates mainly
the $^{18}$Ne ground state (no event to the first excited state in $^{18}$Ne
is observed by SAL). Thus, the measured $^{21}$Na\rpa $^{18}$Ne cross section
can be converted to the $^{18}$Ne($\alpha$,$p_0$)$^{21}$Na$_{\rm{g.s.}}$ cross
section using detailed balance. 

Because of the relatively thick targets that are used in the reverse reaction
experiments, the energy of each data point is not very well defined. This
leads to major uncertainties in the calculation of the reaction rate factor
\Nsv . For example, for the lowest data point of SAL one finds a variation of
the astrophysical \sfact\ of a factor of 5 if the given cross section is
converted to the \sfact\
at the upper or lower energy limit of the data point. Therefore a different
way for the comparison of rate factors from the various experimental
techniques has been chosen in this work (see Fig.~\ref{fig:rate} and
discussion of Eq.~(\ref{eq:exprate}) in Sect.~\ref{sec:MATSAL}).

\subsubsection{Salter {\it et al.}}
\label{sec:sal}
The SAL experiment used the ISAC II facility at TRIUMF.
The energy range under study by SAL for the
\rap\ reaction (from about 1\,MeV up to about
2.6\,MeV) extends the energy range of GRO down to lower energies and reaches
the Gamow window of the $^{18}$Ne\rap $^{21}$Na reaction for the first
time. The SAL data are shown in Fig.~\ref{fig:sigma} as blue points.

From the average cross sections of the ($\alpha$,$p_0$) reaction the
rate factor \Nsv\ is calculated in SAL using the {\sc{exp2rate}} code
\cite{exp2rate}. Because of the missing contributions of the ($\alpha$,$p_{i
  \ne 0}$) channels this \Nsv\ is also considered as a lower limit in
SAL. These missing contributions are estimated in SAL from
Hauser-Feshbach (HF) calculations by T.\ Rauscher.
The calculations indicate that the total rate factor
\Nsv\ is about a factor of three larger than the measured ground-state
contribution. 

The validity of HF calculations is somewhat uncertain in the present
case because of the relatively small level density in the compound nucleus
$^{22}$Mg. However, this uncertainty mainly influences the absolute value of
the calculated cross section which depends on the number of states (or
resonances) at the energy under study. The calculation of the decay branches
in the proton channel to to the ground state and to excited states in
$^{21}$Na should be less affected because it is dominated by the transmission
coefficients. It can also be checked ``by hand'' by calculating
single-particle limits for the proton decay of states in $^{22}$Mg. It is
found that e.g.\ the decay of a hypothetical $0^+$ state in $^{22}$Mg at
$E^\ast = 
9.942$\,MeV ($E = 1.8$\,MeV, i.e.\ in the center of the analyzed energy region
of Fig.~\ref{fig:sigma}) by proton emission proceeds by 31\,\% to the ground
state of $^{21}$Na and by 69\,\% to excited states. This confirms the HF
approach for the decay branch. A further confirmation is obtained from
comparison with the GRO data. Here it is found experimentally that the ground
state branch contributes with 42\,\% to the summed strength.

\subsubsection{ANL data}
\label{sec:ANL}
A similar experiment has also been performed at the Argonne National
Laboratory (ANL). Unfortunately, these data have never been published and can
be found in the ANL Annual Report only \cite{ANL}. The data (extracted from
the figure in \cite{ANL}) are shown in Fig.~\ref{fig:sigma} as red squares.

The ANL data point at the highest energy agrees well with SAL. A second
data point at lower energies is about a factor of two higher, and the error
bars of ANL and SAL are close to overlap. Three upper limits have been
determined at lower energies; these upper limits are slightly higher than the
SAL data points. In total, the unpublished ANL data are in reasonable
agreement with the SAL data. 

As usual, it is very difficult to estimate the
reliability of unpublished data, and thus a publication of the ANL data  would
be very helpful. The ANL data do not enter directly into the recommended rate
(see Sect.~\ref{sec:rate}), but the reasonable agreement with SAL strengthens
the validity of the experimental data by SAL.

\section{Compatibility of results from various experimental techniques}
\label{sec:comp}
\begin{figure}[htb]
\includegraphics[width=\columnwidth,clip=]{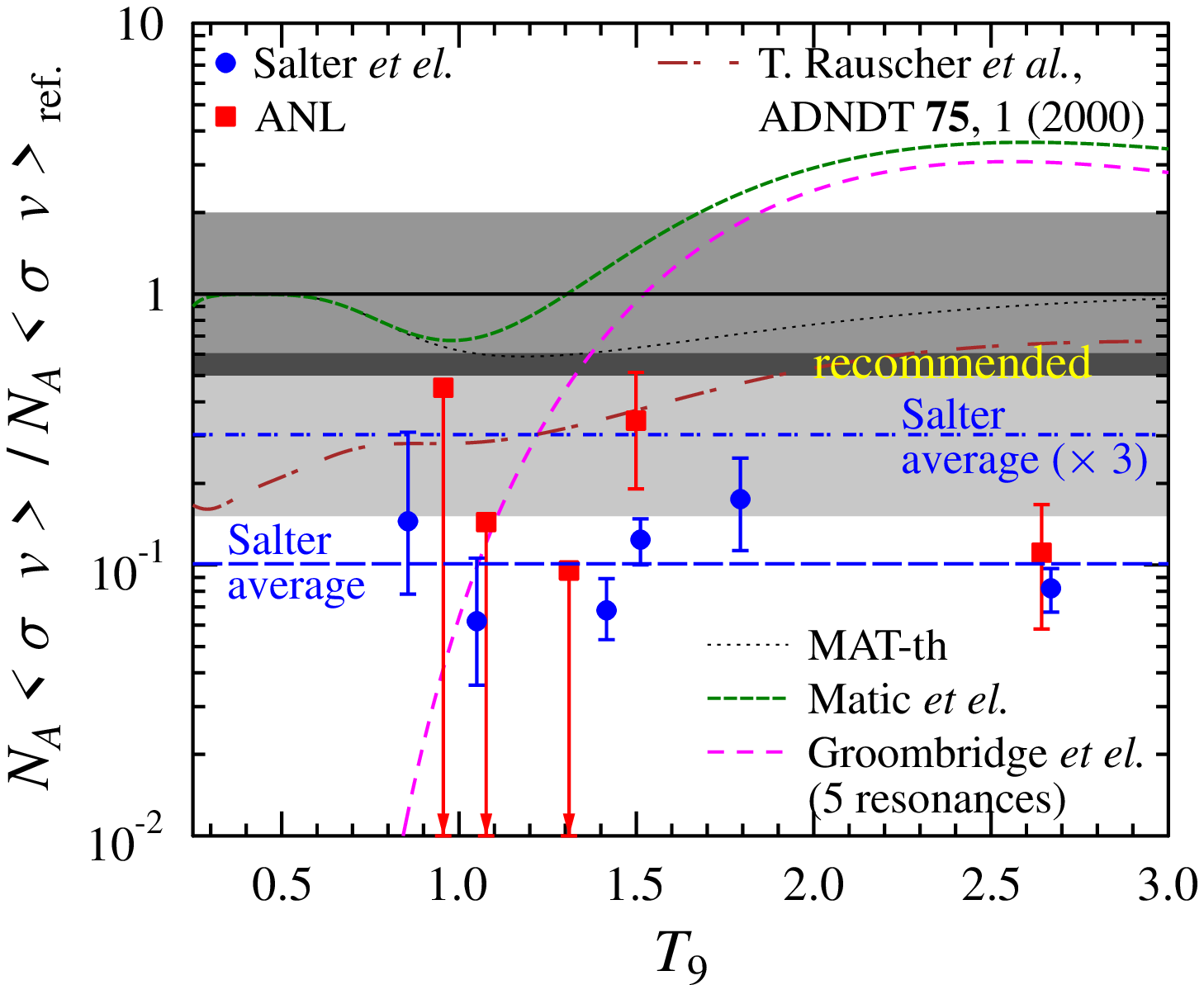}
\caption{
\label{fig:rate}
(Color online)
Ratio between the reaction rate factors \Nsv\ from different studies
normalized to the reference rate factor \Nsv $_{\rm{ref.}}$ from the MAT data
and the modifications in Table \ref{tab:mod}. Long-dashed magenta line: 5
adopted resonances of the GRO experiment; green short-dashed line: original MAT
data; black dotted line: MAT data with calculated strengths $\omega \gamma$
for the 5 resonances of GRO (MAT-th); horizontal blue dashed and dash-dotted
lines: 
average of the SAL data and average multiplied by a factor of three to correct
for the ground-state branching. (Color codes and line styles are identical to
Fig.~\ref{fig:sigma}.) The conversion of the experimental SAL and ANL data to 
the shown \Nsv\ data points is explained in the text. The recommended rate
factor \Nsv $_{\rm{recommended}}$ is located in the narrow overlap of the
error bars of the reverse reaction data (SAL $\times 3$, lightgrey shaded) and
the reference rate (grey shaded) at approx.\ 0.55\,$\times$\,\Nsv
$_{\rm{ref.}}$ (dark grey shaded). Numerical values are listed in Table
\ref{tab:rate}. A theoretical prediction in the statistical model is also
shown (brown dash-dotted) \cite{Rau00}.
}
\end{figure}

In this section we will first analyze whether the experimental data of the
various experimental approaches are compatible with each other. In a second
step the uncertainties of additional ingredients for the calculation of the
rate factor \Nsv\ will be studied. These are in
particular calculated resonance strengths $\omega \gamma$ for the transfer
experiments and the theoretically estimated ground-state branching for the
reverse reaction experiments. The resulting rate factors \Nsv\ are normalized
to the reference factor \Nsv $_{\rm{ref.}}$ determined from the reference
cross section $\sigma_{\rm{ref.}}$, i.e.\ the MAT data with
all revisions shown in Table \ref{tab:mod}. The results are shown in
Fig.~\ref{fig:rate}. For comparison, a theoretical prediction using the
statistical model is also shown \cite{Rau00}.

It has to be pointed out here that a comparison between the average cross
sections $\bar{\sigma}$ in SAL and the resonance data in MAT and GRO is
possible for the first time (see also Fig.~\ref{fig:sigma}) because the total
widths $\Gamma$ of the states in $^{22}$Mg are now available from Table
\ref{tab:width}. 

\subsection{Are the transfer data compatible with the direct data?}
\label{sec:MATGRO}
Here a strict comparison of experimental data is limited to excitation
energies $E^\ast$ or resonance energies $E$ which have been determined in the
MAT transfer and the GRO direct experiment. 
5 of 8 resonances measured by GRO have been uniquely assigned to states which
have been seen in the transfer experiment of MAT. Two further resonances have
not been seen in the MAT transfer experiment, but have been detected in other
transfer experiments (see Table V in MAT). The lowest resonance in GRO is
tentatively assigned to a doublett of states in the MAT data. Thus, there is no
strict contradiction between the experimental transfer data and the direct
data by GRO.

For the calculation of reaction rate factors \Nsv\ from Eq.~(\ref{eq:rate})
MAT have used the strengths of the 5 uniquely assigned resonances.
The remaining 3 resonances of GRO are neglected. However, some
other states have been observed by MAT in the energy range of the GRO
experiment, and theoretical strengths have been used in MAT for these
resonances. Consequently, the calculated \Nsv\ of MAT and GRO agree well
within the temperature range where the observed resonances in GRO define the
rate factor \Nsv\ whereas the \Nsv\ of GRO is much smaller at lower
temperatures (see Fig.~\ref{fig:rate}).

A further compatibility test can be made. The resonance strength $\omega
\gamma$ has a general upper limit: $\omega \gamma \le \omega \Gamma/2$. The
total widths $\Gamma$ vary between 21 and 144\,keV for the resonances in GRO
\cite{Matic07}. These widths impose an experimental 
upper limit on the rate factor \Nsv\ which is about one order of magnitude
above the GRO result. 
The relatively large strengths in GRO are thus fully
compatible, but at the upper limit of the allowed range, especially in the
expected case of $\Gamma_\alpha \ll \Gamma_p \approx \Gamma$. 

This finding is
further strengthened by a theoretical calculation for the only state with a
firm spin assignment. The $2^+$ state at $E = 2130$\,keV has been considered
as a member of a $Q = 10$ higher-nodal rotational band in a simple \al
-cluster model of $^{22}$Mg = $^{18}$Ne $\otimes$ \al\ with the semi-magic
($N=8$) $^{18}$Ne core (similar to \cite{Abe93,Wil02}). Even for
well-established \al -cluster states in the neighboring nuclei
$^{19}$F = $^{15}$N $\otimes$ \al\ and $^{20}$Ne = $^{16}$O $\otimes$
\al\ with their semi-magic ($N=8$) or doubly-magic ($N=Z=8$) cores
it has been found that the calculated width $\Gamma_\alpha$ in this model
overestimates the experimental width by at least 30\,\% and often by about a
factor of two. Here we find for the $2^+$ resonance at 2130\,keV an
experimental strength of $\omega \gamma = 10.3^{+8.6}_{-1.4}$\,keV which
corresponds to $\Gamma_\alpha = 2.06^{+1.72}_{-0.28}$\,keV for $\Gamma_\alpha
\ll \Gamma_p$. The cluster model predicts $\Gamma_\alpha = 1.86$\,keV. Again,
the experimental results of GRO are at the upper limit of the allowed
range. 

The same procedure has been repeated for the other 4 resonances from
GRO which had entered the original calculation of the rate factor in MAT. The
results are listed in Table \ref{tab:theo}. In all cases the \al\ width
derived from the resonance strengths of the GRO experiment are close or even
above the theoretical upper limit.
This is a very unusual finding. However,
because only tentative spin assignments are available for these remaining 4
resonances, it is not possible to strictly exclude the GRO results from the
above theoretical considerations. This holds in particular for the $[4^+]$
resonance at 2.287\,MeV where the GRO width exceeds the theoretical limit by
almost a factor of 10. In this case the reduction factors from Table
\ref{tab:red} suggest $J \le 2$ to allow for the observed resonance strength
of GRO.
\begin{table*}[thb]
\caption{\label{tab:theo}
Properties of 5 states in $^{22}$Mg from the GRO data and comparison of the
derived $\Gamma_\alpha^{\rm{GRO}}$ from $\omega \Gamma_\alpha \approx \omega
\gamma$ to a theoretically estimated maximum width
$\Gamma_\alpha^{\rm{th,max}}$ (discussion see text). All energies $E$ are
given in MeV; widths $\Gamma$ and resonance strengths $\omega \gamma$ are
given in keV. 
}
\begin{tabular}{crr@{$\pm$}lrrr@{$\pm$}lcccc}
\hline
\multicolumn{1}{c}{$E^{\rm{GRO}}$}
& \multicolumn{1}{c}{$E^{\ast\rm{GRO}}$}
& \multicolumn{2}{c}{$\Gamma^{\rm{GRO}}$}
& \multicolumn{1}{c}{$E^{\rm{MAT}}$}
& \multicolumn{1}{c}{$E^{\ast\rm{MAT}}$}
& \multicolumn{2}{c}{$\Gamma^{\rm{MAT}}$}
& \multicolumn{1}{c}{$J^\pi$}
& \multicolumn{1}{c}{$\omega \gamma^{\rm{GRO}}$}
& \multicolumn{1}{c}{$\Gamma_\alpha^{\rm{GRO}}$}
& \multicolumn{1}{c}{$\Gamma_\alpha^{\rm{th,max}}$} \\
\hline
2.17$\pm$0.14 & 10.312 & 130 & 80  & 2.130 & 10.272 &  21 &  3 & $2^+$ 
  & $10.3^{+8.6}_{-1.4}$ & $2.06^{+1.72}_{-0.28}$ & 1.87 \\
2.28$\pm$0.15 & 10.422 & 210 & 100 & 2.287 & 10.429 & 144 & 26 & $[4^+]$ 
  & $ 7.3^{+9.7}_{-1.5}$  & $0.81^{+1.08}_{-0.17}$ & 0.093 \\
2.52$\pm$0.14 & 10.662 & 100 & 50  & 2.509 & 10.651 &  73 & 19 & $[3^-]$ 
  & $18.2^{+8.9}_{-1.9}$ & $2.60^{+1.27}_{-0.27}$ & 1.81 \\
2.72$\pm$0.14 & 10.862 & 210 & 10  & 2.731 & 10.873 &  40 & 12 & $[0^+]$ 
  & $45.2^{+14.6}_{-11.8}$ & $45.2^{+14.6}_{-11.8}$ & 97.1 \\
2.87$\pm$0.14 & 11.012 & 100 & 20  & 2.859 & 11.001 & 136 & 13 & $[4^+]$ 
  & $ 8.1^{+2.9}_{-2.0}$ & $0.90^{+0.32}_{-0.22}$ & 1.23 \\
\hline
\end{tabular}
%
\end{table*}

\subsection{Are the transfer data compatible with the reverse reaction data?}
\label{sec:MATSAL}
A direct comparison between the MAT excitation energies and resonance
strengths and the SAL average cross sections is difficult because there are no
common observables in the different experimental approaches. Nevertheless, a
comparison can be made in the following way. From Eq.~(\ref{eq:av}) and
$\sigma_{\rm{ref.}}(E)$, see Sect.~\ref{sec:mat}, the average cross section
$\bar{\sigma}(\alpha,p)$ can be calculated for each data point of SAL and
ANL. These calculated $\bar{\sigma}(\alpha,p)$ should be about a factor of 
$\approx 3$ larger because the reverse reaction data determine only the
ground-state contribution $(\alpha,p_0)$. The factor of $\approx 3$ is taken
from the HF calculations in SAL (see also discussion in
Sect.~\ref{sec:sal}). The results are listed in Table \ref{tab:av}. 
\begin{table}[tbh]
\caption{\label{tab:av}
Average cross sections $\bar{\sigma}(\alpha,p)$ from $\sigma_{\rm{ref.}}(E)$
compared to the SAL and ANL data. Further discussion see text.
}
\begin{tabular}{r@{$\pm$}lcccrc}
\hline
\multicolumn{2}{c}{$E_{\rm{eff}}(\alpha,p)$}
& \multicolumn{1}{c}{exponent}
& \multicolumn{1}{c}{$\sigma_{\rm{exp}}$}
& Ref.
& \multicolumn{1}{c}{$\bar{\sigma}_{\rm{ref.}}$}
& \multicolumn{1}{c}{$\sigma_{\rm{exp}}/\bar{\sigma}_{\rm{ref.}}$} \\
\multicolumn{2}{c}{(MeV)}
& \multicolumn{1}{c}{for $\sigma$ in mb} \\
\hline
1.194  & 0.130  & $10^{-4}$  & $5.5^{+6.2}_{-3.5}$  
  & SAL  & 37.8 & $0.145^{+0.164}_{-0.067}$ \\
1.379  & 0.129  & $10^{-3}$  & $3.8^{+2.7}_{-1.6}$  
  & SAL  & 61.6 & $0.062^{+0.044}_{-0.026}$ \\
1.683  & 0.121  & $10^{-2}$  & $2.3^{+0.7}_{-0.5}$  
  & SAL  & 33.7 & $0.068^{+0.021}_{-0.015}$ \\
1.758  & 0.069  & $10^{-2}$  & $3.1 \pm 0.6$       
  & SAL  & 25.1 & $0.124 \pm 0.024$ \\
1.970  & 0.117  & $10^{-1}$  & $1.7^{+0.7}_{-0.6}$  
  & SAL  &  9.7 & $0.175^{+0.073}_{-0.062}$ \\
2.568  & 0.061  & $10^{0}$   & $1.7 \pm 0.3$       
  & SAL  & 20.7 & $0.082 \pm 0.015$ \\
1.748  & 0.077  & $10^{-2}$  & $7.5^{+3.8}_{-3.3}$       
  & ANL & 22.0  & $0.341^{+0.173}_{-0.150}$ \\
2.551  & 0.077  & $10^{0}$   & $2.0^{+1.0}_{-0.9}$       
  & ANL & 17.8  & $0.111^{+0.056}_{-0.053}$ \\
\hline
\end{tabular}
%
\end{table}

The ratios $\sigma_{\rm{exp}}/\bar{\sigma}_{\rm{ref.}}$ vary between 0.06 and
0.15 for the SAL data with a geometric mean of 0.101. The higher energy data
point of the ANL data is in good agreement with the corresponding SAL data point
with a ratio of 0.11 compared to 0.08 from SAL. The lower point of ANL is
slightly higher by a factor of about 2.5 but has an uncertainty of a factor of
two. The upper limits of the ANL data are compatible with the SAL data points
at lower energies. Thus, in general the experimental data of SAL and ANL are
in reasonable agreement.

As pointed out above (see Sec.~\ref{sec:inv}), a calculation of reaction rate
factors \Nsv\ from experimental cross sections with relatively large
uncertainties in the energy may have large uncertainties and thus may be
misleading. Here we estimate the rate factors from the reverse reaction
experiments in the following way. We adopt the energy dependence of
$\sigma_{\rm{ref.}}(E)$ and use the ratio
$\sigma_{\rm{exp}}/\bar{\sigma}_{\rm{ref.}}$:
\begin{equation}
N_A <\sigma v>_{\rm{exp}}(T) = N_A <\sigma v>_{\rm{ref.}}(T) \times 
\frac{\sigma_{\rm{exp}}(E)}{\bar{\sigma}_{\rm{ref.}}(E)}
\label{eq:exprate}
\end{equation}
The temperature $T$ for each data point is taken from the most effective
energy $E_{\rm{eff}}$ of the Gamow window which is given by the well-known
relation  $E_{\rm{eff}}/{\rm{keV}} = 122 \times (Z_P^2 Z_T^2 A_{\rm{red}}
T_9^2)^{1/3}$. This leads to the experimental data points shown in
Fig.~\ref{fig:rate}. Their average value is almost exactly a factor of 10
lower than the reference rate \Nsv $_{\rm{ref.}}$ (horizontal blue dashed line
in Fig.~\ref{fig:rate}). Because of the experimental uncertainties of the SAL
and ANL data, a better determination of the temperature dependence of the rate
factor \Nsv\ from experimental data for the reverse reaction is not possible.

A strict comparison of $\bar{\sigma}_{\rm{ref.}}$ and $\sigma_{\rm{exp}}$ requires
two theoretical considerations. First, calculated resonance strengths $\omega
\gamma$ have been used in the calculation of $\sigma_{\rm{ref.}}(E)$, and,
second, the
calculated ground-state branching of about one third (as suggested in SAL, see
also Sect.~\ref{sec:sal}) is responsible for an expected factor of three
discrepancy between $\bar{\sigma}_{\rm{ref.}}$ and
$\sigma_{\rm{exp}}$. However, the ratio 
$\sigma_{\rm{exp}}/\bar{\sigma}_{\rm{ref.}}$ turns out to be about 0.1 (see
Table \ref{tab:av}); i.e., it is a further factor of three smaller than
expected. Both calculations ($\omega \gamma$ for the transfer data, the ground
state branching for the reverse reaction data)
are based on simple but reasonable arguments, and
the uncertainties should not exceed a factor of two. This factor of two for
the uncertainties of \Nsv $_{\rm{ref.}}$ and \Nsv\ from the reverse reaction
data are shown as shaded areas in Fig.~\ref{fig:rate}. Thus, the real rate can
be estimated as follows.

From the transfer data the reference rate \Nsv $_{\rm{ref.}}$ is derived with
an uncertainty of a factor of two; i.e., the real rate should be located in
the interval between $0.5 \, \times$\,\Nsv $_{\rm{ref.}}$ and $2.0 \,
\times$\,\Nsv $_{\rm{ref.}}$. From the reverse reaction the best estimate for
the real rate is $0.3 \, \times$\,\Nsv $_{\rm{ref.}}$ (taking into account the
ground state branching of about $1/3$ as discussed above), again with an
uncertainty of about a factor of two; i.e., the real rate should be located in
the interval between $0.15 \, \times$\,\Nsv $_{\rm{ref.}}$ and $0.6 \,
\times$\,\Nsv $_{\rm{ref.}}$. Combining the above intervals, for the real rate
only a narrow window around $(0.5 - 0.6) \, \times$\,\Nsv $_{\rm{ref.}}$
remains to be compatible with all experimental results. Uncertainties for this
finally recommended rate will be given in Sect.~\ref{sec:rate}.

\subsection{Are the direct data compatible with the reverse reaction data?}
\label{sec:GROSAL}
A strict comparison between the direct GRO data and the reverse reaction data
is possible because GRO have determined the main decay branch of the
resonances seen in their experiment. In particular, according to GRO, a
resonance at $2.52 \pm 0.14$\,MeV with a width of $\Gamma= 100 \pm 50$\,keV
decays mainly to the $p_0$ channel with a resonance strength of $\omega \gamma
= 18.2^{+8.9}_{-1.9}$\,keV. MAT have assigned $J^\pi = [3^-]$, $E = 2.509 \pm
0.013$\,MeV, and $\Gamma = 73 \pm 19$\,keV. The experimental data points at
the highest energies of the SAL and ANL data around 2.5\,MeV are 
affected by this resonance. Using the resonance energy $E$ from MAT, the total
width from the MAT experiment (see Table \ref{tab:width}), and the resonance
strength $\omega \gamma$ from GRO, 
we find $\bar{\sigma} = 40.2$\,mb for the SAL data
point at 2.568\,MeV which is a factor of 24 higher than the experimental value
of $1.7 \pm 0.3$\,mb. A similar factor of 23 is found for the ANL data point
of $2.0^{+1.0}_{-0.9}$\,mb where $\bar{\sigma} = 45.3$\,mb is calculated. A
much better agreement is found if the huge resonance strength of $\omega \gamma
= 18.2$\,keV from the GRO data is replaced by the calculated strength of
$\omega \gamma = 1.12$\,keV (see Table \ref{tab:mod}): 
$\bar{\sigma} = 2.47$\,mb for the SAL data point
and 2.79\,mb for the ANL data point. The role of this resonance turns out to
be minor for the total reaction cross section (see Fig.~\ref{fig:sigma}) which
is dominated by the long tail of the strong $0^+$ resonance at 2.731\,MeV.

This leads to the clear conclusion that there is a strict contradiction
between the experimental data of GRO in the direct experiment and the
experimental SAL data using the reverse reaction. This conclusion is
independent of any theoretical calculations. It seems more likely that there
is an experimental problem in the normalization of the GRO data because
the unpublished ANL data for the reverse reaction are in agreement with the
SAL data, and thus there must be a problem in the two independent SAL and ANL
experiments if the GRO data are correct.

\section{Recommended reaction rate factor \Nsv }
\label{sec:rate}
The above discussion leads to the following recommendations for the reaction
rate factor \Nsv\ of the $^{18}$Ne\rap $^{21}$Na reaction. The most realistic
estimate from the overlap of the uncertainties in Fig.~\ref{fig:rate} is
located around 0.55\,$\times$\,\Nsv $_{\rm{ref.}}$. Consequently, this rate
factor is 
recommended for further use in astrophysical calculations. Numerical values
are listed in Table \ref{tab:rate}.
\begin{table}[tbh]
\caption{\label{tab:rate}
Recommended rate factor \Nsv $_{\rm{recommended}}$ in
cm$^3$\,s$^{-1}$\,mol$^{-1}$ and realistic
lower and upper limits. Note that \Nsv $_{\rm{recommended}}$ =
0.55\,$\times$\,\Nsv $_{\rm{ref.}}$. The lower limit is given by
0.30\,$\times$\,\Nsv $_{\rm{ref.}}$, and the upper limit is given by \Nsv
$_{\rm{ref.}}$.  
}
\begin{tabular}{p{0.8cm}r@{$\times$}lr@{$\times$}lr@{$\times$}l}
\hline
\multicolumn{1}{l}{$T_9$}
& \multicolumn{2}{c}{recommended}
& \multicolumn{2}{c}{lower}
& \multicolumn{2}{c}{upper} \\
\hline
   0.1  & 4.1 & $10^{-24}$  & 2.3 & $10^{-24}$  & 7.5 & $10^{-24}$ \\
   0.2  & 1.7 & $10^{-15}$  & 9.6 & $10^{-16}$  & 3.2 & $10^{-15}$ \\
   0.3  & 3.0 & $10^{-11}$  & 1.6 & $10^{-11}$  & 5.4 & $10^{-11}$ \\
   0.4  & 1.1 & $10^{-08}$  & 6.1 & $10^{-09}$  & 2.0 & $10^{-08}$ \\
   0.5  & 7.2 & $10^{-07}$  & 4.0 & $10^{-07}$  & 1.3 & $10^{-06}$ \\
   0.6  & 1.7 & $10^{-05}$  & 9.3 & $10^{-06}$  & 3.1 & $10^{-05}$ \\
   0.7  & 2.1 & $10^{-04}$  & 1.1 & $10^{-04}$  & 3.8 & $10^{-04}$ \\
   0.8  & 1.7 & $10^{-03}$  & 9.3 & $10^{-04}$  & 3.1 & $10^{-03}$ \\
   0.9  & 1.0 & $10^{-02}$  & 5.7 & $10^{-03}$  & 1.9 & $10^{-02}$ \\
   1.0  & 4.8 & $10^{-02}$  & 2.7 & $10^{-02}$  & 8.8 & $10^{-02}$ \\
   1.1  & 1.8 & $10^{-01}$  & 1.0 & $10^{-01}$  & 3.3 & $10^{-01}$ \\
   1.2  & 5.7 & $10^{-01}$  & 3.2 & $10^{-01}$  & 1.0 & $10^{+00}$ \\
   1.3  & 1.5 & $10^{+00}$  & 8.5 & $10^{-01}$  & 2.8 & $10^{+00}$ \\
   1.4  & 3.7 & $10^{+00}$  & 2.0 & $10^{+00}$  & 6.7 & $10^{+00}$ \\
   1.5  & 8.0 & $10^{+00}$  & 4.4 & $10^{+00}$  & 1.5 & $10^{+01}$ \\
   1.6  & 1.6 & $10^{+01}$  & 8.9 & $10^{+00}$  & 2.9 & $10^{+01}$ \\
   1.7  & 3.0 & $10^{+01}$  & 1.7 & $10^{+01}$  & 5.5 & $10^{+01}$ \\
   1.8  & 5.4 & $10^{+01}$  & 3.0 & $10^{+01}$  & 9.8 & $10^{+01}$ \\
   1.9  & 9.1 & $10^{+01}$  & 5.0 & $10^{+01}$  & 1.7 & $10^{+02}$ \\
   2.0  & 1.5 & $10^{+02}$  & 8.3 & $10^{+01}$  & 2.7 & $10^{+02}$ \\
   2.1  & 2.4 & $10^{+02}$  & 1.3 & $10^{+02}$  & 4.3 & $10^{+02}$ \\
   2.2  & 3.7 & $10^{+02}$  & 2.0 & $10^{+02}$  & 6.7 & $10^{+02}$ \\
   2.3  & 5.6 & $10^{+02}$  & 3.1 & $10^{+02}$  & 1.0 & $10^{+03}$ \\
   2.4  & 8.2 & $10^{+02}$  & 4.5 & $10^{+02}$  & 1.5 & $10^{+03}$ \\
   2.5  & 1.2 & $10^{+03}$  & 6.6 & $10^{+02}$  & 2.2 & $10^{+03}$ \\
   2.6  & 1.7 & $10^{+03}$  & 9.3 & $10^{+02}$  & 3.1 & $10^{+03}$ \\
   2.7  & 2.4 & $10^{+03}$  & 1.3 & $10^{+03}$  & 4.3 & $10^{+03}$ \\
   2.8  & 3.3 & $10^{+03}$  & 1.8 & $10^{+03}$  & 6.0 & $10^{+03}$ \\
   2.9  & 4.5 & $10^{+03}$  & 2.5 & $10^{+03}$  & 8.1 & $10^{+03}$ \\
   3.0  & 6.0 & $10^{+03}$  & 3.3 & $10^{+03}$  & 1.1 & $10^{+04}$ \\
\hline
\end{tabular}
%
\end{table}

Uncertainties for \Nsv\ may be estimated as follows. A realistic lower limit
can be taken from the SAL data (multiplied by a factor of three to take into
account the ground-state branching) which is shown as a blue dash-dotted line
in Fig.~\ref{fig:rate}. A realistic upper limit is the reference rate factor
\Nsv $_{\rm{ref.}}$. A strict lower limit provide the SAL data (without the
correction of the ground-state branching, see blue dashed line in
Fig.~\ref{fig:rate}). A strict upper limit in the astrophyically most
relevant temperature range is about a factor of three higher
than the reference rate factor. It can be taken from the GRO results; although
these results seem to be questionable, it has been shown that the GRO
strengths are close to theoretical upper limits and thus suitable to provide
an upper limit.

Finally, this leads to a recommended reaction rate factor \Nsv
$_{\rm{recommended}}$ = 
0.55\,$\times$\,\Nsv $_{\rm{ref.}}$ with a realistic uncertainty of a factor
of 1.8 and an extreme uncertainty of a factor of 5.5; it is interesting to
note that after all the estimates of uncertainties one ends up with an almost
Gaussian uncertainty distribution with a factor of 1.8 for a realistic (1
sigma) uncertainty and a factor of about 5.5 for an extreme uncertainty (3
sigma).

Of course, the reduction factor of 0.55 between the reference calculations and
the final recommended rate factor also has to be applied to the reference
cross section $\sigma_{\rm{ref.}}$ shown in Fig.~\ref{fig:sigma}. The
absolute value of $\sigma_{\rm{ref.}}$ depends on the calculated resonance
strengths $\omega \gamma$ which are proportional to the decay widths
$\Gamma_\alpha$ into the \al\ channel (for $\Gamma_\alpha \ll
\Gamma_p$). Thus, all calculated $\Gamma_\alpha$ of MAT and in Table
\ref{tab:mod} should be reduced by the same factor of 0.55.

The recommended rate factor \Nsv $_{\rm{recommended}}$ is slightly lower than
the MAT rate factor around $T_9 \approx 1$, but significantly smaller at
higher temperatures around $T_9 \approx 2$. The minor difference between the
MAT rate factor and \Nsv $_{\rm{recommended}}$ at lower
temperatures is due to a compensation of the enhancement from new spin
assignments from the CHA transfer data (leading to smaller spins $J$ and thus
increased resonance strengths) and the derived reduction factor of 0.55 from
the comparison with the reverse reaction data. The significant decrease at
higher temperatures is mainly a consequence of the replacement of the huge
resonance strengths from the GRO experiment by smaller calculated resonance
strengths. 

In a comparison of the new recommended rate factor \Nsv $_{\rm{recommended}}$
with the SAL result (see Table II in \cite{Sal12}) it has to be kept in mind
that SAL provide the rate factor for the ground-state contribution
$^{18}$Ne(\al ,$p_0$)$^{21}$Na$_{\rm{g.s.}}$ which is derived from their
experimental data; the given upper and lower limits
are calculated from their experimental uncertainties, but do not include the
additional contributions of the (\al ,$p_{i \ne 0}$) channels and the
corresponding uncertainties. Thus, it is not
surprising that the new recommended rate factor exceeds the upper limit of the
SAL rate factor. The new recommended rate factor is about a factor of 5
higher than the SAL result in the astrophysically most relevant temperature
range of $T_9 = 1 - 2$.

The theoretical prediction of the $^{18}$Ne\rap $^{21}$Na cross section in
\cite{Rau00} is based on the statistical model and a global parameter set. The
calculation has been done before the experimental results of GRO, SAL, MAT,
ANL, CHA, and MAT were available. By definition, such a statistical model
calculation cannot 
reproduce details of the $\sigma(E)$ curve shown in Fig.~\ref{fig:sigma}.
Nevertheless, the predicted rate factor \Nsv\ is in reasonable agreement with
the recommended rate factor (see Fig.~\ref{fig:rate}) and remains within the
realistic uncertainty estimate of \Nsv $_{\rm{recommended}}$ in the
temperature range $T_9 = 1 - 3$. However, the temperature dependence of the
theoretical rate factor is slightly steeper compared to the recommended rate
factor, and thus at low temperatures below $T_9 \approx 1$ the theoretical
rate factor is located below the recommended rate factor between the realistic
and the the extreme lower limit of the recommendation.

The recommended rate factor \Nsv $_{\rm{recommended}}$ is fitted by the usual
expression, see e.g.\ \cite{Rau00}, Eq.~(16):
\begin{eqnarray}
\frac{N_A < \sigma v >}{{\rm{cm}}^3 {\rm{s}}^{-1} {\rm{mol}}^{-1}} = 
  & & \, \, \exp{(a_0 + a_1 T_9^{-1} + a_2 T_9^{-1/3} + a_3 T_9^{1/3}} 
\nonumber \\ 
  & & \, \, \, \,
  + a_4 T_9 + a_5 T_9^{5/3} + a_6 \ln{T_9})
\label{eq:fit}
\end{eqnarray}
The $a_i$ parameters are listed in Table \ref{tab:fit}. The deviation of the
fitted rate factor is always below 10\,\% over the full temperature range $0.25
\le T_9 \le 3$ and typically below 5\,\% in the most relevant range $1 \le T_9
\le 2$.
\begin{table*}[tbh]
\caption{\label{tab:fit}
Fit parameters $a_i$ of the reaction rate factor \Nsv\ from
Eq.~(\ref{eq:fit}).
}
\begin{tabular}{ccccccc}
\hline
\multicolumn{1}{c}{$a_0$}
& \multicolumn{1}{c}{$a_1$} 
& \multicolumn{1}{c}{$a_2$} 
& \multicolumn{1}{c}{$a_3$} 
& \multicolumn{1}{c}{$a_4$} 
& \multicolumn{1}{c}{$a_5$} 
& \multicolumn{1}{c}{$a_6$} \\
\hline
  -21.0595
& -0.3301
& -58.1167
& 89.3359
& -14.8713
& 1.9862
& -24.1080 \\
\hline
\end{tabular}
%
\end{table*}

\section{Summary and conclusions}
\label{sec:summ}
The present knowledge of the reaction rate factor \Nsv\ of the $^{18}$Ne\rap
$^{21}$Na reaction has been summarized. For this purpose experimental results
from different experimental techniques are combined. Transfer reactions
provide the best determination of the excitation energy $E^\ast$ and spin and
parity $J^\pi$ of states in $^{22}$Mg which appear as resonances in the
$^{18}$Ne\rap $^{21}$Na reaction; however, transfer reactions cannot provide
the required resonance strengths $\omega \gamma$. These strengths have to be
taken from theory or from direct experiments which are however 
extremely difficult and require the combination of a radioactive $^{18}$Ne
beam and a helium gas target. Complementary information has been derived from
the experimental study of the reverse $^{21}$Na\rpa $^{18}$Ne reaction using a
radioactive $^{21}$Na beam and a solid CH$_2$ target.

A basic prerequisite for the
comparison of results from various experimental techniques is the availability
of total widths $\Gamma$ for the resonances under study. The total widths
$\Gamma$ were determined from a reanalysis of the peak widths in the MAT
experiment.

A compatibility test between the results from various 
experimental techniques shows that
there is no contradiction between the various
experimental data except the disagreement
between the direct GRO data and the reverse reaction data from SAL and
ANL. This leads to the conclusion that the most likely explanation is a
problem in the normalization of the GRO data. Consequently, resonance
strengths from GRO have been replaced by theoretical resonance strengths in
the calculation of the rate factor \Nsv . 

The calculation of \Nsv\ for the $^{18}$Ne\rap $^{21}$Na reaction from transfer
data requires theoretical resonance strengths, and the calculation of
\Nsv\ from the reverse $^{21}$Na\rpa $^{18}$Ne reaction data requires
a theoretical estimate of the (\al ,$p_0$) ground-state branching. Both
calculations are based on simple but reasonable arguments, and the
corresponding uncertainties should not exceed a factor of two. This leads to a
relatively narrow overlap region between the higher \Nsv\ calculated from
transfer and the lower \Nsv\ calculated from the reverse reaction data. This
narrow overlap region is considered as the new 
recommended reaction rate factor \Nsv $_{\rm{recommended}}$. The uncertainty
of the recommended rate factor is about a factor of 1.8 ($\approx 1\,\sigma$
uncertainty). For $T_9 = 1 - 3$ a theoretical prediction \cite{Rau00} lies
within this error band, but the theoretical temperature dependence of the rate
factor \Nsv\ is somewhat steeper than the new recommendation.

The new recommended rate factor is slightly lower than the MAT rate factor at
low temperatures and significantly smaller at higher temperatures, and the new
rate factor exceeds the SAL result by about a factor of 5.
The strong conclusion of SAL (based on their lower limit for the rate factor
\Nsv ) that 
``the breakout from the HCNO cycle via the $^{18}$Ne\rap $^{21}$Na reaction is
delayed and occurs at higher temperatures than previously predicted'' cannot
be supported. Instead, because of the only minor deviations of \Nsv
$_{\rm{recommended}}$ from the MAT result at low temperatures around $T_9 =
1$, the earlier conclusions of MAT should remain valid
in general. Further astrophysical network calculations with the new
recommended rate factor \Nsv $_{\rm{recommended}}$ are required to
study the relevance of the modified temperature dependence of the rate factor
in detail.

\acknowledgments
We thank M.\ Aliotta, A.\ M.\ van den Berg, G.\ P.\ A.\ Berg,
K.-E.\ Rehm, P.\ Salter, M.\ Wiescher for encouraging discussions,
and T.\ Rauscher for his code {\sc{exp2rate}}. This work
was supported by OTKA (NN83261).

\end{document}